\documentclass[aapm,mph,print,graphicx]{revtex4-1} 
\usepackage{graphicx}
\usepackage{amssymb}
\usepackage{amsmath}
\usepackage{siunitx}
\usepackage{verbatim} 
\usepackage{color}
\usepackage{setspace} 
\usepackage{soul} 
\usepackage[mathlines]{lineno}
\begin{document}
\title{Dynamic Block Matching to assess the longitudinal component of the dense motion field of the carotid artery wall in B-mode ultrasound sequences -- Association with coronary artery disease \\ \vspace{10pt} 
{\color{blue}Article published in Medical Physics 45(11), 5041--5053, 2018\\
https://doi.org/10.1002/mp.13186}}
\author{Guillaume Zahnd}
\altaffiliation{
Imaging-based Computational Biomedicine Lab,\\
Nara Institute of Science and Technology,\\
8916-5 Takayama-cho, Ikoma, Nara 630-0192, Japan\\
Computer Aided Medical Procedures,\\
Technische Universit{\"a}t M{\"u}nchen,\\
Boltzmannstra{\ss}e 3, 85748 Garching, Germany\\
g.zahnd@tum.de\\
}
\author{Kozue Saito}
\altaffiliation{
Department of Stroke and Cerebrovascular Diseases,\\
National Cerebral and Cardiovascular Center,\\
5-7-1 Fujishiro-dai, Suita, Osaka 565-8565, Japan\\
}
\author{Kazuyuki Nagatsuka}
\altaffiliation{
Department of Stroke and Cerebrovascular Diseases,\\
National Cerebral and Cardiovascular Center,\\
5-7-1 Fujishiro-dai, Suita, Osaka 565-8565, Japan\\
}
\author{Yoshito Otake}
\altaffiliation{
Imaging-based Computational Biomedicine Lab,\\
Nara Institute of Science and Technology,\\
8916-5 Takayama-cho, Ikoma, Nara 630-0192, Japan\\
}
\author{Yoshinobu Sato}
\altaffiliation{
Imaging-based Computational Biomedicine Lab,\\
Nara Institute of Science and Technology,\\
8916-5 Takayama-cho, Ikoma, Nara 630-0192, Japan\\
}
\date{\today}
\begin{abstract}
\begin{description}
\item[Purpose] 
The motion of the common carotid artery tissue layers along the vessel axis during the cardiac cycle, observed in ultrasound imaging, is associated with the presence of established cardiovascular risk factors. However, the vast majority of the (semi-)automatic methods devised to measure this so-called \textit{``longitudinal kinetics''} phenomenon are based on the tracking of a single point, thus failing to capture the overall---and potentially inhomogeneous---motion of the entire arterial wall. The aim of this work is to introduce a motion tracking framework able to simultaneously extract the temporal trajectory of a large collection of points (several hundred) horizontally aligned and spanning the entire exploitable width of the image, thus providing a dense motion field.
\item[Method]
The only action required from the user is to indicate the left and right borders of the region to be processed. A previously validated contour segmentation method is used to position one point in the arterial wall in each column of the image. Between two consecutive frames, the radial motion of each point is predetermined by the position of the segmentation contours. The longitudinal motion, which is the main focus of the present work, is determined in two steps. First, a series of independent block matching operations are carried out for all the tracked points. Here, the displacement of each point is not determined yet, instead the similarity map is stored. Then, an original dynamic-programming approach is exploited to regularize the collection of similarity maps and estimate the globally optimal motion over the entire vessel wall. Sixty-two atherosclerotic participants at high cardiovascular risk were involved in this study. Method training and validation was performed with 20 and 42 participants, respectively. The amplitude-independent index $\sigma X$ was introduced to quantify the motion inhomogeneity across the length of the artery.
\item[Results] 
A dense displacement field, describing the longitudinal motion of the carotid far wall over time, was extracted from all participants. For each cine-loop, the method was evaluated against manual reference tracings performed on three local points, and showed a good accuracy, with an average absolute error $(\pm~\text{STD})$ of $150~(\pm 163)~$\SI{}{\micro\metre}. It also demonstrated an overall greater robustness compared to a previously validated method based on single-point motion tracking. For all the 62 participants, the analyzed region had, in average, a width of \SI{24.2}{\milli\metre}, involving the simultaneous tracking of 357~points along 151 temporal frames, and requiring a total computational time of \SI{68}{\second}. Analyzing the inhomogeneity of the carotid artery motion showed a strong correlation between $\sigma X$ and the presence of coronary artery disease ($\beta\text{-coefficient}=.586$, $p=.003$).
\item[Conclusions] 
To the best of our knowledge, this is the first time that a method is specifically proposed to assess the dense motion field corresponding to the longitudinal kinetics of the carotid far wall. This approach has potential to evaluate the homogeneity (or lack thereof) of the wall dynamics. The proposed method has promising performances to improve the analysis of arterial longitudinal motion and the understanding of the underlying patho-physiological parameters.
\item[Keywords] Ultrasound; Carotid Artery; Motion Tracking; Dynamic Programming; Cardiovascular Risk Factors
\end{description}
\end{abstract}
\maketitle
\section{Introduction}\label{intro}
\begin{figure}[h]
\begin{center}
\includegraphics[width=.7\textwidth]{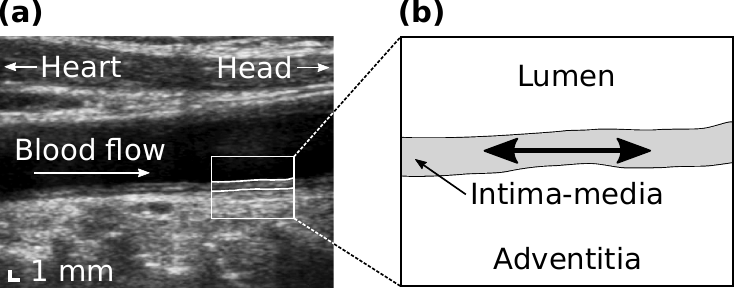}
\caption{\label{fig_loki}
(a)~Longitudinal view of a common carotid artery in B-mode ultrasound.
(b)~Detail of the far wall. The cyclic motion of the tissue layers along the axis of the vessel (\emph{i.e.},~LOKI) is indicated by the double arrow.
}
\end{center}
\end{figure}

Cardiovascular risk evaluation is a major public health issue as well as a tremendous scientific challenge. In the last recent years, the characterization of the elastic deformation of the tissue layers of the common carotid artery along the direction parallel to the vessel axis during the cardiac cycle in ultrasound sequences (also called cine-loops) has gathered a growing attention. This patho-physiological phenomenon, hereafter dubbed as \textit{LOKI} for \textit{longitudinal kinetics}, corresponds to the shear between the intima-media complex and the tunica adventitia (Figure~\ref{fig_loki})~\cite{persson2003new}. This motion was shown to be cyclic and reproducible over several months~\cite{ahlgren2012different}. 
\par
Although the causes---as well as the implications---of LOKI remain to be fully determined, this dynamic parameter has been demonstrated by several groups to be associated with the presence of established cardiovascular risk factors.
LOKI was shown to be associated with carotid atherosclerotic plaque burden in man and mouse~\cite{svedlund2011alongitudinal}, and to be a predictor for 1-year cardiovascular outcome in patients with coronary artery disease~\cite{svedlund2011carotid}.
The motion amplitude was found to be significantly reduced in elderly diabetic patients compared to young healthy subjects~\cite{zahnd2011measurement}. 
It was also confirmed that LOKI was significantly reduced in patients with periodontal disease compared to age-matched healthy volunteers, independently of other established cardiovascular risk
factors~\cite{zahnd2012longitudinal}.
Similar results were found in a study involving individuals with spinal cord injury and able-bodied subjects, where the longitudinal retrograde motion and retrograde shear strain were reduced in the subclinical cardiovascular risk population~\cite{tat2017reduced}.
Administration of catecholamines and $\beta\text{-blockade}$ in pigs showed a significant positive correlation with LOKI, thus potentially constituting a link between mental stress and cardiovascular disease~\cite{ahlgren2012longitudinal}.
Building upon these findings, the independence between LOKI and wall shear stress was demonstrated, suggesting that LOKI is most likely caused by the pulsatile displacement of the heart, rather than by blood flow~\cite{ahlgren2015profound}.
A similar hypothesis was formulated in a study that evaluated the association of LOKI with both left ventricular cardiac motion and local blood velocity~\cite{au2016carotid}.
A strong association between LOKI and the severity of carotid stenosis was demonstrated, potentially constituting a tool to screen arterial sites known to be predisposed to atherosclerotic plaque formation~\cite{soleimani2015novel}.
In addition to motion amplitude, two new indices of arterial stiffness derived from LOKI waveforms were explored: the total distance of the intima trajectory during one heart cycle, as well as the degree of a polynomial function fitting LOKI, were found to be associated with arterial stiffness measurements~\cite{yli2016new}.
It was further demonstrated that LOKI is inversely associated with arterial stiffening~\cite{taivainen2017interrelationships}, and that LOKI provides complementary information regarding arteriosclerosis and risk factors compared to traditional markers~\cite{taivainen2018influence}.
These findings therefore strongly suggest that robust and accurate LOKI quantification may improve cardiovascular risk evaluation.
\par
Tracking the motion of the common carotid artery \emph{in vivo} in B-mode ultrasound image cine-loops is hindered by several factors of different nature. 
First, the resolution cell of the scanner is generally coarser in the direction perpendicular to the beam ($x$) than in the direction of the beam propagation ($y$). This is reflected by the anisotropic shape of the point spread function, whose width (determined by the probe geometry and the depth of the focal zone) is typically larger than its height (determined by the ultrasound signal wave-length)~\cite{meunier1995ultrasonic}. 
Second, the profile of the image in the longitudinal direction is rather homogeneous and does not present clear landmarks. As opposed to the radial direction, where image intensity undergoes profound variations along the depth of the image, only a small gray level variation is present along the longitudinal profile, as visible in Figure~\ref{fig_loki}. This phenomenon is known as the aperture problem, and in this case, is caused by the geometry of the anatomical structure of the vessel, consisting of tissue layers aligned along the longitudinal axis. As a result, motion along the $x$ direction is more difficult to perceive.
Third, images are often corrupted by several degrading phenomena inherent to the ultrasound modality itself, such as out-of-plane movements, speckle decorrelation, acoustic shadowing, or movement artifacts, which can lead to the alteration of the tracked speckle pattern over time. Another aspect to consider, especially in the context of clinical routine, is the high variability of image quality caused by the specific vessel geometry and tissue echogenicity of each examined subject.
\par
A number of motion tracking methods have been put forward to address the aforementioned issues and evaluate LOKI. 
A block matching approach---also called speckle tracking---could successfully extract the longitudinal trajectory of a selected point in a given cine-loop~\cite{golemati2003carotid}. The motion was assessed between pairs of consecutive frames by locating the point centered on the speckle pattern bearing the greatest similarity with the tracked reference pattern, according to the sum of absolute difference metric. This study however has two limitations: First, the rather large size of the tracked blocks ($3.2 \times 2.5)~\text{mm}^2$, added to the fact that the images were not interpolated, resulted in coarse, stairway-like motion trajectories. Second, this method was applied on 5 healthy subjects and 2 atherosclerotic patients, but the tracking accuracy was not quantitatively evaluated against any reference. Another approach exploited a very small block size ($0.5 \times 0.5)~\text{mm}^2$ and reached a high tracking accuracy when evaluated \emph{in vitro} in an agar phantom with tiny glass beads producing distinct echoes~\cite{cinthio2005evaluation}. Reference motion was provided using laser measurements and the mean tracking error was \SI{5.9}{\micro\metre}. Nevertheless, \emph{in vivo} experiments were only conducted on one healthy subject and not evaluated quantitatively. Moreover, this framework requires optimal image quality and a careful selection of a single and very well defined scatter point, thus making it not truly applicable on images acquired in routine clinical practice. A different approach was proposed based on cross-correlation after compensation for luminance changes~\cite{yli2013axial}. Applied in 19 healthy subjects, the evaluation protocol did not involve characterizing the accuracy of the method \emph{per se}, but rather focused on quantifying the intra-subject reproducibility of the measurements on image data acquired on two subsequent days, which resulted in a Cronbach $\alpha$ coefficient between $-0.68$ and $0.93$. Several advanced methods based on Kalman filtering have been specifically introduced in an attempt to improve the tracking performances. An approach predicting the motion of the target point and the appearance of the corresponding speckle pattern was put forward~\cite{gastounioti2013carotid}. Although quantitative evaluation was only performed on simulated sequences, showing a 47\% accuracy increase compared to traditional block matching, the method was applied on 9 healthy subjects and 31 atherosclerotic patients, demonstrating substantial difference in the measured motion amplitude.
Another approach proposed the use of a control signal designed to avoid the accumulation of successive tracking errors~\cite{zahnd2013evaluation}. Quantitatively evaluated in 57 healthy subjects and 25 patients at high cardiovascular risk against manually traced motion references, the average accuracy ($\pm~\text{STD}$) was $84~(\pm 107)$~\SI{}{\micro\meter}. A statistically significant difference was also found in the motion amplitude between controls and patients ($643 \pm 274$~\SI{}{\micro\meter} \emph{vs} $408 \pm 281$~\SI{}{\micro\meter}, $p<0.0001$). The combination of update and prediction has also been exploited within a state-space approach based on an elasticity-model and ruled by a $H_\infty$ filter~\cite{gao2017robust}. A quantitative evaluation of the tracking accuracy on 37 healthy subjects and 103 patients at high cardiovascular risk was performed against manually tracked points, the Pearson's correlation coefficient was $0.9536$ and the width of the $95\%$ confidence interval was \SI{387}{\micro\meter}.
\par
All the methods mentioned above share a similar property: They are based on the local estimation of the temporal trajectory of a single point. This approach requires a careful selection of the best point to be processed within the image, typically, a point located in a region with optimal image quality, and centered on a well distinguishable speckle pattern. Such local analysis is limited by three principal drawbacks: First, interrogation of a unique region fails to capture the non-uniform behaviour of the dynamics of the wall, since different regions may undergo deformations of different magnitudes~\cite{zahnd2015progressive}. Second, tracking a single target point is likely to yield erroneous results in noisy image regions. Third, point selection---and hence results exploitation---is subject to inter-analyst variability. Therefore, characterizing the motion of the entire wall instead of a single point, which is the aim of the present method, would be a more advantageous approach to address these issues.
\par
A few methods only have been proposed to assess the global deformation of the arterial tissues across an elongated region of interest (ROI) covering (part of) the full width of the image. The traditional block matching framework has been extended by independently tracking a small set of points and averaging the resulting trajectories~\cite{zahnd2012longitudinal,tat2017reduced}. These simple models, however, do not exploit any assumption on the physical interrelations between the motion of the different points. Individual tracking failures, although their magnitude is attenuated by the averaging operation, are therefore likely to have an impact on the resulting overall trajectory. The Velocity Vector Imaging commercial software (VVI, Research Arena 2; TomTec imaging systems GmbH, Unterschleissheim, Germany) has been applied for LOKI quantification~\cite{svedlund2011longitudinal}. However, that tool was originally designed to measure the heart dynamics in echocardiography, and the evaluation of its performances on the carotid artery showed poor results~\cite{zahnd2013evaluation}. Another approach based on feature detectors and descriptors to automatically identify and track several keypoints was recently introduced~\cite{scaramuzzino2017longitudinal}. 
However, three limitations can be observed: First, due to substantial computation time, this method relies on a limited amount of points (typically, 10) and therefore only captures a rather sparse distribution of the wall dynamics. 
Second, the $(x,y)$ coordinates of the reference points to be tracked must be reselected at each time step, with no guarantee that the same set of points will be continuously analyzed over time. Therefore the motion trajectory of a specific sub-region over time can not be captured.
Third, for each pair of consecutive frames, a single displacement value is computed with the median of the motion resulting from all the tracked points. This yields a unique trajectory that describes the global displacement of the entire wall, but does not permit to study the specific displacement of different sub-regions of the wall individually.
\par
The aim of this study is to propose an approach capable of assessing the dense motion field of the tissue layers, namely, tracking one point in every column of the image. The rationale is an attempt to minimize individual tracking errors by exploiting the collaboration between the ensemble of points. This is achieved by \emph{simultaneously} extracting, in each frame of the cine-loop, the displacement of all the points by means of combinatorial analysis. More precisely, the methodological contribution of the present study is the following. The framework is based on two main steps. First, for each point (one per column), a block matching operation is conducted, exploiting three specific features (radial motion guidance with anatomical contour segmentation; use of the initial speckle pattern as a control signal to generate the tracked reference block; storage of the similarity matching describing the ensemble of potential displacements). Second, the matching score of all points are gathered in an artificial space to generate a single cost function. Then, a dynamic programming scheme is run to determine the minimal-cost path that describes the globally optimal displacement of all the points simultaneously. The framework is applied to \emph{in vivo} data, validated against manual reference annotations, and assessed in the context of a pilot clinical study.
\section{Materials and Methods}\label{sec_methods}
\begin{figure}[h]
\begin{center}
\includegraphics[width=.45\textwidth]{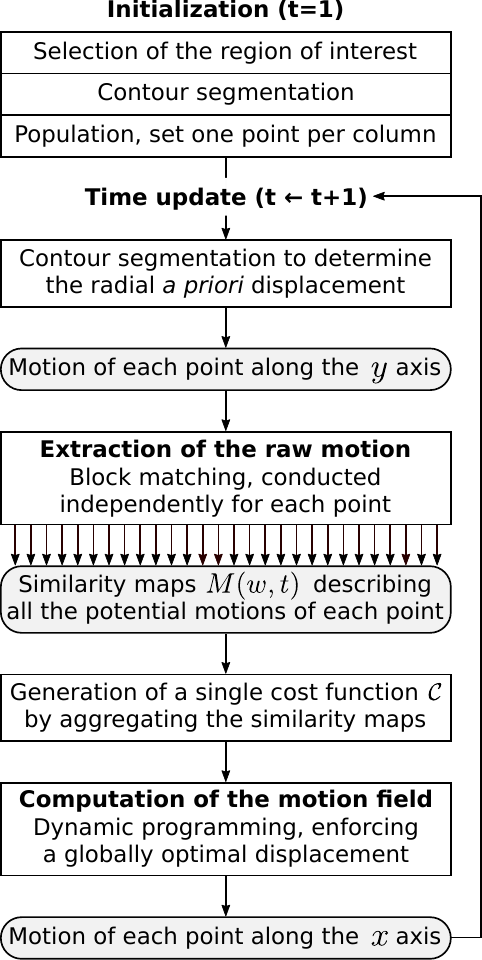}
\caption{\label{fig_diagram}
Illustration of the main steps of the framework.
}
\end{center}
\end{figure}

The method introduced here is devised to simultaneously extract the motion of all the points located on a line. The main steps of the method---hereafter referred to as Dynamic Block Matching (DBM)---are illustrated in Figure~\ref{fig_diagram}. The framework is composed of two principal operations: First, a block matching search is conducted independently for all the assessed points, then the resulting similarity maps are exploited in the context of a dynamic programming scheme to determine the global optimal displacement map of the ensemble of points. The remainder of this section is organized as follow: First the image acquisition procedure is defined, then the DBM framework is presented in detail, next the evaluation protocol is described, finally a quantitative analysis between LOKI-derived parameters and cardiovascular risk factors is proposed.
\subsection{Data acquisition and study sample}
Ultrasound image acquisition was performed at the National Cerebral and Cardiovascular Center, Osaka, Japan, by a skillful medical doctor (KS), as part of the routine procedure, using a clinical scanner (Toshiba Applio, Tokyo, Japan) equipped with a linear probe (PLT-704SBT, \SI{7.5}{\mega\hertz}). Pixel size was \SI{70}{\micro\metre} and frame rate was 32~fps. Images were recorded in the longitudinal plane during several cardiac cycles (average number of frames in each cine-loop: $151 \pm 44$, corresponding to $4.7 \pm$\SI{1.4}{\second}). The probe was oriented such that the left and right sides of the image corresponded to the heart and head direction, respectively.
\par
Sixty-two participants suffering from atherosclerosis (age: $76 \pm 7$ y.o., 50 male) were enrolled in this study. The participants characteristics is provided in Table~\ref{tab_patients}. Opt-out consent was obtained from all participants. This observational study was conducted as a retrospective analysis and fulfilled the requirements of our institutional review board and the ethics committee.
\begin{table}[h!]
\caption{Characteristics of the 62 involved participants.}
\label{tab_patients}
\begin{center}
\begin{tabular}{llll}
\hline
Parameter & Unit & Total & Missing \\ 
\hline
Coronary artery disease  & $n~(\%)$ & 21 (34\%) & -- \\
Hypertension              & $n~(\%)$ & 53 (85\%) & -- \\
Systolic blood pressure   & mm Hg & $129 \pm 15$ & 14 \\
Diastolic blood pressure  & mm Hg & $71 \pm 9$ & 14 \\
Pulse pressure            & mm Hg & $58 \pm 15$ & 14 \\
Dyslipidemia      & $n~(\%)$ & 55 (89\%) & -- \\
Peak systolic value       & m/s   & $92 \pm 67$ & 13 \\
Other heart diseases      & $n~(\%)$ & 27 (44\%) & -- \\
Diabetes mellitus      & $n~(\%)$ & 24 (39\%) & -- \\
Gender                    & male, $n~(\%)$ & 50 (81\%) & -- \\
Age                       & years & $76 \pm 7$ & 1 \\
Previous stroke           & $n~(\%)$ & 15 (24\%) & -- \\
Intima-media thickness    & mm & $1.59 \pm 0.86$ & 8 \\
Peripheral artery disease & $n~(\%)$ & 8 (16\%) & -- \\
Estimated glomerular filtration rate & mL/min/1.73m\textsuperscript{2} & $54.9 \pm 20.4$ & 16 \\
Total cholesterol         & mmol/L & $157 \pm 29$ & 24 \\
Glycosylated hemoglobin & \% & $6.7 \pm 1.0$ & 29 \\
Smoking:                  & $n~(\%)$ && 4 \\
- Current                 & & 7 (11\%)  & \\
- Ex                      & & 39 (63\%) & \\
- Never                   & & 12 (19\%) & \\
Body mass index           & kg/m\textsuperscript{2} & $22.9 \pm 3.2$ & 8 \\
Other vascular diseases   & $n~(\%)$ & 6 (10\%) & -- \\
\hline
\end{tabular}
\end{center}
\end{table}
\subsection{Motion tracking}
\subsubsection{Initialization}
\begin{figure}[h!]
\begin{center}
\includegraphics[width=\textwidth]{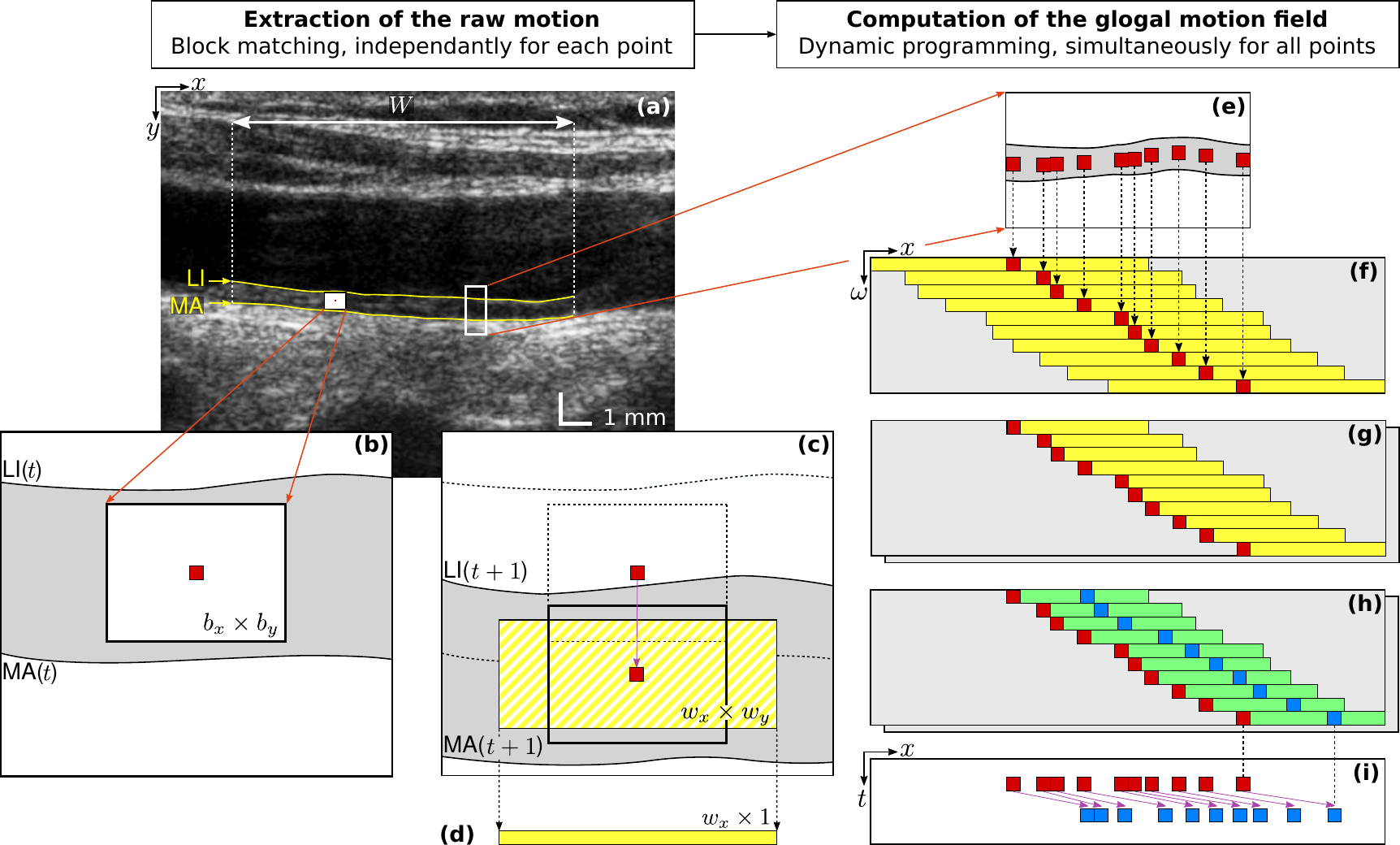}
\caption{\label{fig_framework}
Schematic representation of the main steps of the framework.
(a)~Carotid ultrasound image. The lumen-intima (LI) and media-adventitia (MA) contours are segmented (yellow solid lines) within the region of interest of width $W$ (white dashed lines).
The white-filled rectangle represents a block $\mathcal{B}(\omega, t)$ (size $b_x \times b_y$), centered around a point $p(\omega, t)$ to be tracked.
(b)~Detail of the intima-media complex (gray region), with the tracked block (white rectangle) centered around $p(\omega, t)$ (red point, size $1 \times 1$ pixel).
(c)~The search window (yellow rectangle, size $w_x \times w_y = 71 \times 3$~pixels in the interpolated image), yielding a 2D SSD map $M(\omega, t)$ (same yellow rectangle), is automatically positioned in the center of the intima-media complex in the next $(t+1)^{st}$ frame. The radial displacement of $p(\omega, t)$ is indicated by the vertical arrow.
(d)~Similarity vector $V(\omega, t)$ (size $w_x \times 1$), generated from the column-wise minimum of the 2D SSD map $M(\omega, t)$ in panel c.
(e)~Detail of a sub-sample of the $\Omega$ tracked points (red points).
(f)~Cost function~$\mathcal{C}$, resulting from the aggregation of the individual similarity~vectors (panel d).
Here (in f-h), the $\omega$ axis corresponds to the left-to-right indices of the points.
(g)~Cost functions~$\mathcal{C}^+$ (front) and~$\mathcal{C}^-$ (back), enforcing a unidirectional motion field (towards positive or negative values of $x$, respectively).
(h)~Result of the front propagation in $\mathbb{C}^\pm$ ruled by Equation~\ref{eq_propag}, showing the optimal path $\mathcal{O}^\pm$ (blue points). A more detailed representation of this panel is presented in Figure~\ref{fig_propagation}.
(i)~Motion field $\mathcal{X}$ resulting from the optimal path (arrows between the red and blue points), yielding the points coordinate in the $(t+1)^{st}$ frame (in this example, $\mathcal{O} = \mathcal{O}^+$).
}
\end{center}
\end{figure}

First, a ROI is determined by manually indicating the left and right borders of the full exploitable width $W$ of the far wall in order to clip out noisy regions (Figure~\ref{fig_framework}a). All the subsequent steps of the framework are fully-automatic.
\par
In each frame $t$ of the image cine-loop, the contours of the lumen-intima (LI) and media-adventitia (MA) interfaces are segmented within the ROI (Figure~\ref{fig_framework}a), using a previously validated method~\cite{zahnd2017fully}. Briefly, the contour segmentation framework is based on a front propagation approach specifically devised to simultaneously (as opposed to one after the other) extract the two anatomical interfaces of the intima-media complex. This is achieved by building a 3D space (LI depth $\times$ MA depth $\times$ width), where the minimum-cost path is a medial axis that describes, in each column of the image, \emph{i)}~the center of the intima-media complex, and \emph{ii)}~its thickness. The LI and MA contours can immediately be deduced from the medial axis.
\par
In preparation for a block matching scheme, the contour information is used to populate a mesh of $\Omega$ points $p(\omega, t)$, with $\omega \in \{1,\dots, \Omega\}$ and $t=1$, in the first frame of the cine-loop. One point is placed in each column of the ROI, in the center of the intima-media complex (Figure~\ref{fig_framework}b,e). 
\par
The image is then spatially interpolated: It has been shown that LOKI magnitude was approximately \SI{1}{\milli\metre}~\cite{cinthio2005evaluation,golemati2003carotid,zahnd2011measurement}. Given an isotropic pixel size of~\SI{70}{\micro\metre}, the total amplitude of the tissue motion is around 14~pixels over the cardiac cycle, which generally corresponds to 30~images. Although tissue dynamics greatly depends on the cardiac phase, the motion amplitude between two consecutive frames generally corresponds to a few pixels and may often be smaller than one pixel. Therefore, to capture the motion in finer details, a spline interpolation with a factor of $10$ is performed along the $x$ direction for sub-pixel resolution, as previously proposed by other studies~\cite{cinthio2005evaluation,zahnd2011measurement,gastounioti2013carotid,tat2017reduced,gao2017robust}. This operation is applied to the image, as well as the points and contour coordinates.
\par
\begin{singlespacing}
\end{singlespacing}
\subsubsection{Extraction of the raw motion}
This operation consists in estimating the motion field $\mathcal{X}(\omega, t)$, namely the relative $x$-wise displacement of each point $p(\omega, t)$ in each frame $t$ of the cine-loop.
This process being carried out independently for each point, the description of the operation will be given for a single point at a given time step.
The previously described interpolation scheme is applied to the frames $t$ and $t+1$. A block matching operation is then carried out between these two consecutive frames: a rectangular block $\mathcal{B}_t$ of size $(b_x \times b_y)$ is centered around~$p(\omega, t)$ in the $t^{th}$~frame, and a search window of size $(b_x \times b_y)$ is explored in the $(t+1)^{st}$~frame. Matching similarity is evaluated with the sum of squared differences (SSD) metric.
\par
Let us briefly recall three specificities that are most generally followed by traditional block matching approaches:
\emph{i)}~the next position of the tracked point $p(\omega, t+1)$ is blindly determined by the coordinates of the global minimum in the SSD map,
\emph{ii)}~the search window in the $(t+1)^{st}$~frame is centered around $p(\omega, t)$, and
\emph{iii)}~the reference block $\tilde{\mathcal{B}_t}$ corresponds to the speckle pattern centered around the currently tracked point $\mathcal{B}_t$.
In comparison with traditional implementations, these three specific points are addressed differently by the present framework, as described hereafter.
\par
First, the search is guided along the $y$ direction using a previously validated contour segmentation method~\cite{zahnd2017fully}. The rationale is to perform a first estimate of the wall radial displacement to guide motion tracking and systematically ensure that the tracked points remain within the intima-media complex. Guidance is realized by automatically setting the $y$ coordinate of the center of the search window in the center of the intima-media complex in the $(t+1)^{st}$ frame (Figure~\ref{fig_framework}c). The segmentation method has been reported to delineate the LI and MA interfaces with a mean absolute error ($\pm~\text{STD}$) of $47(\pm 70)$~\SI{}{\micro\meter} and $55(\pm 68)$~\SI{}{\micro\meter}, respectively~\cite{zahnd2017fully}. Therefore, block matching is conducted along the $y$ direction with a reduced maximum displacement $w_y=0.2~\text{mm}$, in order to allow a fine search around the initial position provided by contour segmentation. In contrast, since no \emph{a priori} guidance is provided along the $x$ direction, a wider exploration range is permitted, with $w_x=0.5~\text{mm}$. This value has been selected based on the maximum expected displacement between two time steps and validated in a previous study~\cite{zahnd2013evaluation}. Given a pixel size of \SI{70}{\micro\meter} and an interpolation factor of $(10, 1)$ in the longitudinal and radial directions, respectively, the size $(w_x \times w_y)$ of the search window in pixels is equal to $(71 \times 3)$.
\par
Second, the reference block $\tilde{\mathcal{B}}$ used within the block matching operation is systematically updated to follow the gray-level variations of the moving images over time, while still preserving the appearance of the initial tracked pattern. The motivation of this implementation is to avoid a potential and irreversible divergence of the trajectory that may be caused by the accumulation of tracking errors due to artifacts and speckle decorrelation. This is accomplished by using a control signal $\mathcal{B}_1$, corresponding to the initial image pattern in the first frame of the cine-loop, to systematically keep the memory of the initially tracked point. To cope with small gray-level variations, the reference block $\tilde{\mathcal{B}_t}$ is thus generated by a weighted sum of the current speckle pattern $\mathcal{B}_t$ and the control signal $\mathcal{B}_1$, according to the following relation:
\begin{equation}
\label{eq_bm}
\tilde{\mathcal{B}_t} = (1-\sigma) \mathcal{B}_t + \sigma \mathcal{B}_{1}
\end{equation}
Here, $\sigma \in [0,1]$ is a weighting factor that determines the relative influence of the initial and current speckle patterns in generating the reference block.
\par
Third, the optimal position of the ensemble of all $\Omega$ blocks is determined simultaneously by means of a dynamic programming scheme. Please note that the aim of this paragraph is to describe the operations undertaken to prepare the dynamic programming scheme, whereas the dynamic programming algorithm \emph{per se} is described thereafter in Section~\ref{dp}. For a given tracked point $p(\omega, t)$, as part of the block matching operation, a search window is explored, resulting in a 2D SSD map~$M(\omega, t)$ of size $w_x \times w_y$ (Figure~\ref{fig_framework}c). From this map, a 1D similarity vector $V(\omega, t)$ of length $w_x$ is generated by selecting the smallest value in each column of the initial 2D SSD map (Figure~\ref{fig_framework}d). Therefore, at this stage, no decision is taken regarding determining the displacement of any single point: instead of selecting the candidate point with the lowest value in the 2D SSD map, the matching potential along the $x$ direction is stored for further use. The rationale is to simultaneously determine the optimal $x$-wise displacement of the ensemble of all blocks by using the collection of similarity vectors in a dynamic programming scheme, as described in the following paragraph. Let us also remember that the displacement along the $y$ axis has already been determined by the contour segmentation.
\subsubsection{Computation of the motion field}\label{dp}
In this step, at a given time $t$, the ensemble of 1D similarity vectors $V(\omega, t)$, resulting from the raw motion estimation of the $\Omega$ points $p(\omega, t)$, are considered collectively. A combinatorial analysis is performed to determine, among all possible solutions describing the displacement of the ensemble of all points, the optimal motion field given the following rules: 
\emph{i)~data similarity:} displacement is guided by the SSD matching criterion represented by the 1D similarity vectors $V(\omega, t)$, 
\emph{ii)~non-crossing trajectories:} the initial ordering of the points along the $x$ axis does not change,
\emph{iii)~motion smoothness:} increase or decrease of the $x$-wise distance between two neighboring points is penalized, and 
\emph{iv)~motion uniformity:} between two successive frames, all moving points must either follow a displacement towards the right or left side of the image.
The implementation enforcing these rules is described below.
\par
A dynamic programming algorithm based on front propagation is proposed. A cost function~$\mathcal{C}$ is constructed in the artificial space $(x, \omega)$, where each $\omega\text{-th}$ 1D similarity vector $V(\omega, t)$ is oriented along the $x$ direction and centered on $\big( \mathcal{X}(\omega, t), \omega \big)$ (Figure~\ref{fig_framework}f).
Here, the $\omega$ axis of the coordinate system corresponds to the left-to-right ranking of the points: two neighbors points $p(\omega_n, t)$ and $p(\omega_{n+1}, t)$ yield $\omega_{n+1} = \omega_n +1$.
The values of~$\mathcal{C}$ outside of the region covered by the maps correspond to non-reachable positions, and are therefore set to infinity.
\par
The condition of motion uniformity is addressed by using $\mathcal{C}$ to generate two cost functions~$\mathcal{C}^\pm$ (Figure~\ref{fig_framework}g). In each row $\omega$ of $\mathcal{C}^+$ (respectively, $\mathcal{C}^-$), the value of the points whose $x$-coordinates are strictly smaller (respectively, larger) than $\mathcal{X}(\omega, t)$ is set to infinity, thus enforcing a global displacement to the right (respectively, left).
\par
A front propagation is then run to build two cumulative cost functions $\mathbb{C}^+$ and $\mathbb{C}^-$. For $\omega=1$, $\mathbb{C}^\pm$ is initialized with the corresponding values of $\mathcal{C}^\pm$. Then, for $\omega \in \{2, \dots, \Omega\}$, $\mathbb{C}^\pm$ is iteratively generated according to Equation~\ref{eq_propag}.
\par
\begin{equation}
\label{eq_propag}
\begin{split}
\mathbb{C}^\pm(x, \omega+1)=\min_{d_x \in \{ -\text{D}_x, \cdots, -1 \}} \Big\{ \mathbb{C}^\pm(x+d_x, \omega) \\
+ \Big( \big(\mathcal{C}^\pm(x, \omega +1)\big)^\gamma + \big(\mathcal{C}^\pm(x+d_x, \omega)\big)^\gamma \vphantom{d_x^{\beta}} \Big) \\
\times  \big( 1+ \alpha |\hat{d} + d_x|^{\beta} \big) \Big\}
\end{split}
\end{equation}
\par
\begin{figure}[h]
\begin{center}
\includegraphics[width=.7\textwidth]{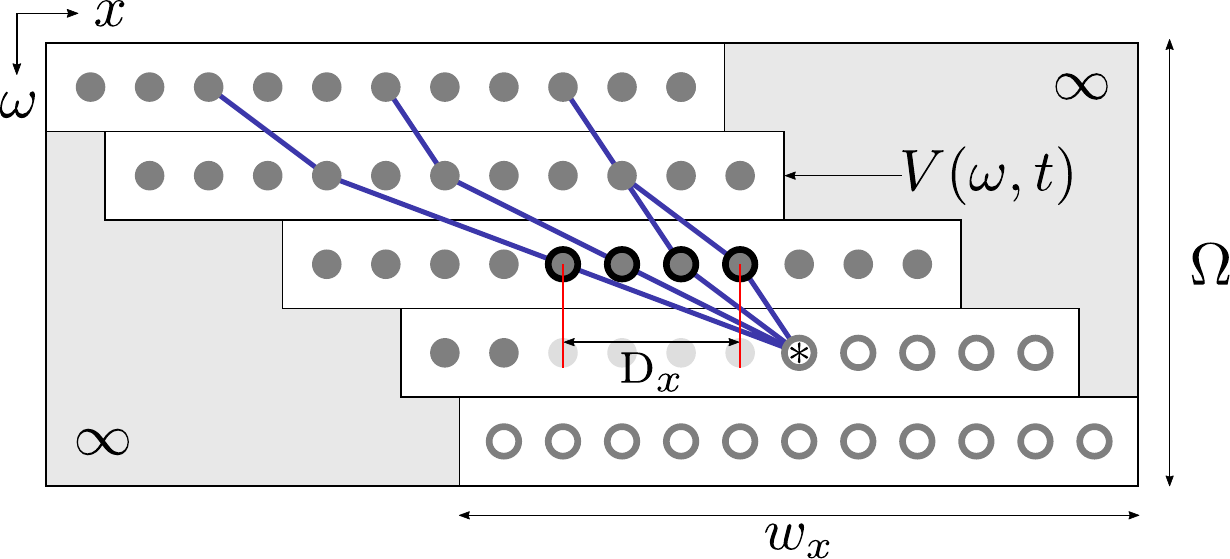}
\caption{\label{fig_propagation}
Front-propagation scheme used to generate the cumulative cost functions $\mathbb{C}^\pm$ from the $\Omega$ similarity vectors~$V(\omega, t)$. For the sake of clarity, this example depicts a non-interpolated image ($\hat{d}=1$). Here, the direction of propagation is top-to-bottom. The gray-filled nodes depict the regions that already have been computed, and the white-filled nodes represent the regions yet to be computed. The node that is currently processed is indicated by an asterisk, and its potential neighbors are circled in black.
In this example, $\Omega=5$ (total number of tracked points), $\text{D}_x=4$ (reachable neighborhood), $w_x=11$ (width of the block matching window).
The black lines connecting the nodes represent the successive back-tracking steps from any given node to the top of the map. Every path connecting the dots corresponds to the potential positioning of a set of points $p_\omega$ arranged along strictly increasing $x$ values. The final solution is determined as the path with the minimal cumulative cost.
}
\end{center}
\end{figure}

To provide a more intuitive grasp of this relation, the cumulated cost of the current node (first line, left part) is determined as the minimal value, across a reachable neighborhood $d_x$, of the addition of \emph{i)} the cumulated cost of the previous node (first line, right part), and \emph{ii)} the sum of the cost of the current and previous nodes (second line) multiplied by a factor to penalize non-realistic distances between the two nodes (third line). Here, $d_x \in \{ -\text{D}_x, \cdots, -1 \}$ prevents crossing point trajectories, namely to respect the condition $\mathcal{X}(\omega_1, t) < \mathcal{X}(\omega_2, t)$ given two points $\omega_1$ and $\omega_2$ such that $\omega_1 < \omega_2$. The elasticity of the mesh is controlled by the smoothing coefficients $\alpha$, $\beta$, and $\gamma$, as well as by the parameter $\hat{d}=10$, which represents the expected $x$-wise distance between two adjacent points, namely one pixel in the original image corresponding to ten pixels in the interpolated image. The distance penalty $|\hat{d} + d_x|$ is null when two neighbors points are separated by exactly ten pixels (one pixel in the non-interpolated image), and gradually becomes greater when the distance between these two points increases (stretching) or decreases (compression). A schematic representation of the front propagation is provided in Figure~\ref{fig_propagation}.
\par
Finally, the optimal paths $\mathcal{O}^+$ and $\mathcal{O}^-$ are extracted in both $\mathbb{C}^+$ and $\mathbb{C}^-$ via back-tracking~\cite{zahnd2017fully} from the point $(x, \Omega)$ having the minimal cumulated cost (Figure~\ref{fig_framework}h). The solution $\mathcal{O}$ with the minimal cost among $\mathcal{O}^+$ and $\mathcal{O}^-$ is used to determine the motion field $\mathcal{X}$ such that $\mathcal{X}(\omega, t+1) = \mathcal{O}(\omega), \omega \in \{ 1, \dots, \Omega \}$ (Figure~\ref{fig_framework}i).
\subsection{Method evaluation}
To evaluate the performance of the method on a fair basis, a training set and a testing set were generated. The training set was composed of 20 randomly selected participants, and was used during the development phase of the method to empirically determine the optimal parameter settings. The testing set was composed of the 42 remaining participants, and was used solely during the evaluation phase of the method, using the previously determined optimal parameter settings, to determine the performance of the proposed approach on independent test data. 
\par
In order to quantify the accuracy of the method despite the lack of ground truth, manual annotations were performed to constitute a reference. For each cine-loop, a set of three reference points $\bar{p}_i,~i=1,2,3$, were first manually placed by an experienced analyst (GZ) in the left, center, and right part of the ROI in the first frame. These anchor points corresponded to well contrasted and distinguishable speckle patterns that remained visible through the entire duration of the cine-loop. Then, the 2D trajectories of these points were manually traced during the entire cine-loop by tracking their position in all frames of the sequence. Intra-analyst variability was also assessed for all tracked points in the testing set. Manual annotations were conducted using a graphical interface that was specifically developed for this purpose, and a Wacom Intuos Pro pen tablet for improved precision.
\par
The DBM method was then applied to the entire ROI. The three columns specifically describing the motion of the reference points $\bar{p}_i$ were selected for evaluation. Finally, a state-of-the-art tracking method based on Kalman Block Matching (KBM)~\cite{zahnd2013evaluation} was used for comparison purpose and applied to track the same set of reference points $\bar{p}_i$. The tracking accuracy, for both the DBM and KBM methods, was evaluated by means of a point-to-point distance comparison between the resulting estimated trajectory and the reference manual tracings. 
\subsection{Statistical analysis}\label{sec_stat}
The peak-to-peak amplitude $\Delta X$ of the motion trajectory (Figure~\ref{fig_traj}c), an established LOKI-derived parameter that was demonstrated to be associated with cardiovascular risk factors~\cite{zahnd2011measurement, zahnd2012longitudinal, tat2017reduced, ahlgren2012longitudinal, ahlgren2015profound, au2016carotid}, was extracted from the resulting motion field (Figure~\ref{fig_traj}b,c). For each participant, $\Delta X$ was automatically determined as the average between all the local peak-to-peak amplitude values, in all the available cardiac cycles and across the entire width of the ROI.
\par
The intra-participant variability $\delta X$ index was introduced to evaluate the reliability of measurements based on a single point, namely to quantify LOKI reproducibility with respect to the selection of the horizontal coordinate of the initial measurement point. This parameter was assessed by means of the following procedure. For each participant, the local amplitude $\Delta X_i$ was automatically measured in each $i^{th}$ column of the ROI. The intra-participant variability $\delta X$ was then established by calculating the mean absolute difference between all pairs $\{\Delta X_a, \Delta X_b\}$, with $a\in [1,\cdots,\Omega-1]$, $b\in [2,\cdots,\Omega]$, $a<b$, and $\Omega$ the total width of the ROI.
\par
The inhomogeneity index $\sigma X$ was introduced to quantify the consistency (or lack thereof) of the motion field across the horizontal direction of the ROI (Figure~\ref{fig_traj}b). The parameter $\sigma X$ is determined as follow: A linear normalization is first applied to the motion field to scale the values between~0 and~1. The aim of this step is to generate an index that is independent of the mere amplitude of the motion. Then $\sigma X$ is obtained by calculating the average standard deviation of the normalized motion field along the horizontal direction. The purpose of this step is to detect, in any given frame of the cine-loop, the presence of spatial regions where LOKI magnitude is greater than in other regions. Such behaviour, caused by an inhomogeneous motion field, is thus reflected by a large $\sigma X$ value.
\par
Linear regression analysis was performed to evaluate the association of participants characteristics (Table~\ref{tab_patients}, predictor variables) with either $\sigma X$ or $\Delta X$ (response variables).
The analysis was performed separately for each response variable.
A univariate model was first generated.
A multivariable model was then generated, with a selection scheme of predictor variables based on three criteria:
\textit{i)}~association to the response variable with $p<0.15$, as derived from the univariate model, 
\textit{ii)}~no pairwise correlation between the predictor variables (in case of two candidates variables being correlated, the one with the strongest association to the response variable was selected), and
\textit{iii)}~adoption of the \emph{``one in twenty''} rule~\cite{austin2015number}, leading to the selection of a total of three predictor variables.
All statistical analysis was undertaken with R~\cite{r2008language}.
The value $p<0.05$ was considered to indicate a statistically significant difference.
\section{Results}\label{sec_results}
\subsection{Parameter settings}
For all the 62 involved cine-loops, the average width $W$ of the processed region, corresponding the to full exploitable width of the image, was $24.2 \pm$\SI{7.0}{\milli\metre}. This corresponded to an average number of tracked points $p_\omega$, defined by the number of columns within the ROI, of $\Omega = 357 \pm 120$. The average distance between two neighboring points $\bar{p}_i$ used to generate the manual references was $7.2 \pm$\SI{4.3}{\milli\metre}.
\par
The optimal value for each parameter was empirically determined using the training set ($n=20$) by successively running the method using different parameters values and selecting the configuration that yielded the lowest tracking error. The allowed range of parameters is detailed in Table~\ref{tab_params}. The method was then applied once to the testing set ($n=42$). The following parameter settings were used: size of the block $\mathcal{B}$, $(b_x \times b_y) = 0.65 \times 0.50~\text{mm}^2$; size of the search window, $(w_x \times w_y) = 0.50 \times 0.20~\text{mm}^2$; weight of the control signal, $\sigma=0.35$; smoothing coefficients, $\alpha = 0.5$, $\beta = 2.0$, $\gamma=0.5$.
\begin{table}[h!]
\caption{Allowed range of the parameter settings during the empirical training phase. Parameters with more tentative values are those having the largest impact on the performances.}
\label{tab_params}
\begin{center}
\begin{tabular}{c|l}
\hline
$b_x$ (\SI{}{\micro\metre}) & 0.15, 0.25, 0.30, 0.35, 0.40, 0.50, 0.55, 0.60, 0.65, 0.70\\
$\sigma$ & 0.15, 0.25, 0.35, 0.75\\ 
$\gamma$ & 0.10, 0.25, 0.50, 1.00\\
$\alpha$ & 0.10, 0.25, 0.50\\
$\beta$ & 1.00, 2.00\\
\hline
\end{tabular}
\end{center}
\end{table}
\subsection{Tracking performances}\label{sec_rob}
\begin{figure}[h!]
\begin{center}
\includegraphics[width=1\textwidth]{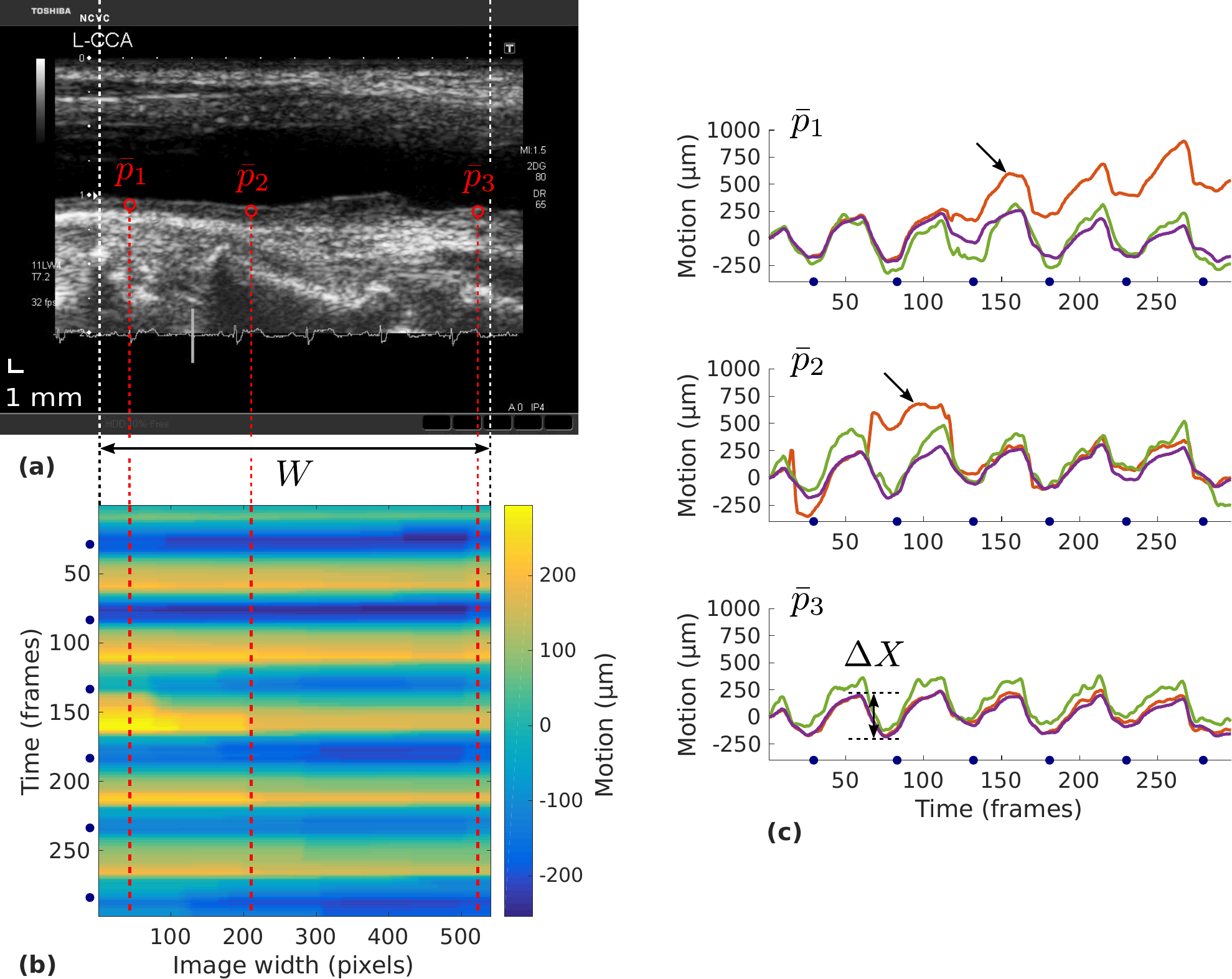}
\caption{\label{fig_traj}
Representative example result from the testing set.
(a)~Image depicting the region of interest of width $W$ (ROI, white dashes) and the set of three points $\bar{p}_i$ used for evaluation (red circles).
(b)~Dense LOKI field $\mathcal{X}$ (2D) resulting from the DBM method, displaying the longitudinal motion across the entire width of the ROI. The red dashes indicate the trajectory of the points $\bar{p}_i$.
(c)~Trajectories (1D) of the three evaluated points, for the DBM method (purple), the manual reference tracings (green), and the KBM method (orange). The tracking failures of the KBM method are indicated by the arrows. The peak-to-peak amplitude $\Delta X$ of the motion trajectory is represented. The blue dots on the temporal axes (b,c) represent the end-diastole.
}
\end{center}
\end{figure}

\begin{figure}[h]
\begin{center}
\includegraphics[width=1\textwidth]{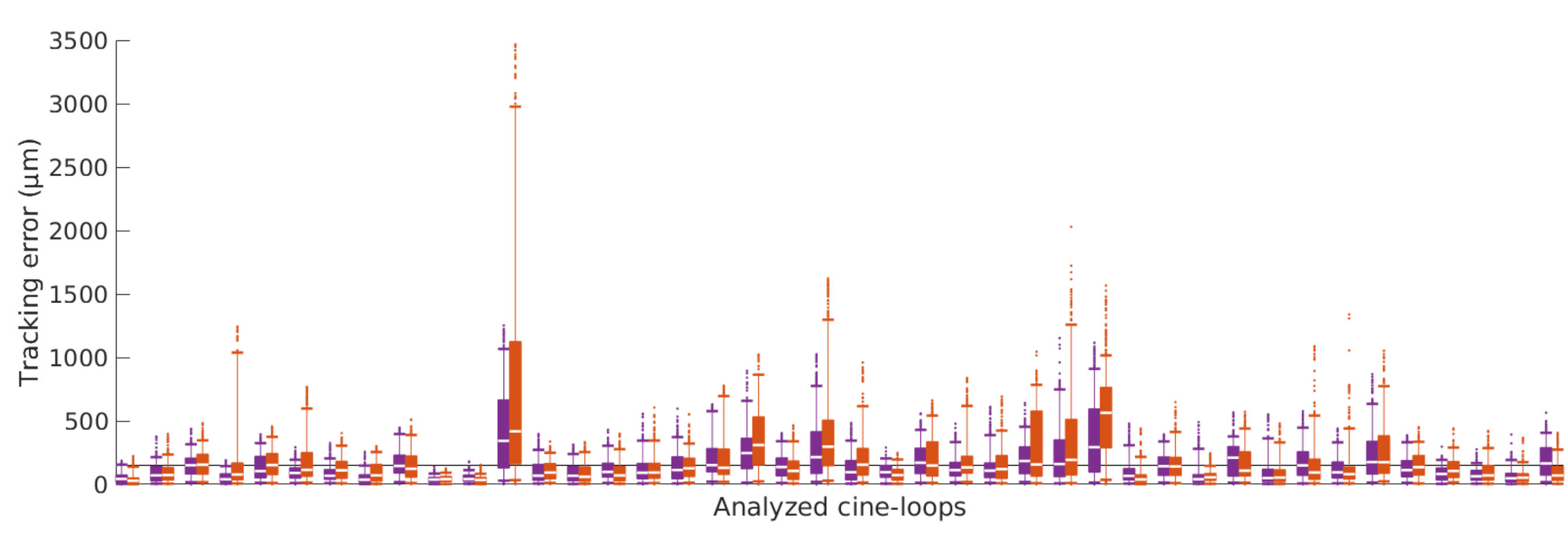}
\caption{\label{fig_boxplot}
Illustration of the absolute tracking errors, for all the 42 participants of the testing base, for the DBM (purple) and KBM (orange) methods. Percentiles are indicated by boxes (25th and 75th), inner lines (50th) and error bars (5th and 95th). The \SI{150}{\micro\metre} level is represented by the black horizontal line.
}
\end{center}
\end{figure}

The motion was successfully extracted from all the analyzed cine-loops. The quantitative evaluation of the proposed DBM framework, showing the mean absolute ($\pm~\text{STD}$) difference between the trajectories resulting from the manual reference annotations and those resulting from the method, is presented in Table~\ref{tab_results}, together with the comparison against the previously introduced KBM method. Overall, the temporal trajectories resulting from the DBM method demonstrated a good similarity with the manual reference annotations, as displayed in Figure~\ref{fig_traj}c. It was also observed that the DBM method performed quite robustly, as opposed to the KBM method, which in some cases failed to capture the exact tissue motion due to progressive drift or sudden jumps to another target point. This is reflected by the Figure~\ref{fig_boxplot}, where outliers are more numerous in the case of the KBM method.
\begin{table}[h!]
\caption{Average absolute tracking error ($\pm$ standard deviation) in \SI{}{\micro\metre} compared against manual reference tracings for the Dynamic Block Matching (DBM) and Kalman Block Matching (KBM) methods, and intra-analyst variability (MAN).}
\label{tab_results}
\begin{center}
\begin{tabular}{ccc}
\hline
Method & Training set ($n=20$) & Testing set ($n=42$) \\
DBM    & $141 \pm 118$ & $150 \pm 163$ \\
KBM    & $180 \pm 248$ & $191 \pm 270$ \\
MAN    & -- & $129 \pm 146$ \\
\hline
\end{tabular}
\end{center}
\end{table}
\par
\begin{figure}[h!]
\begin{center}
\includegraphics[width=1\textwidth]{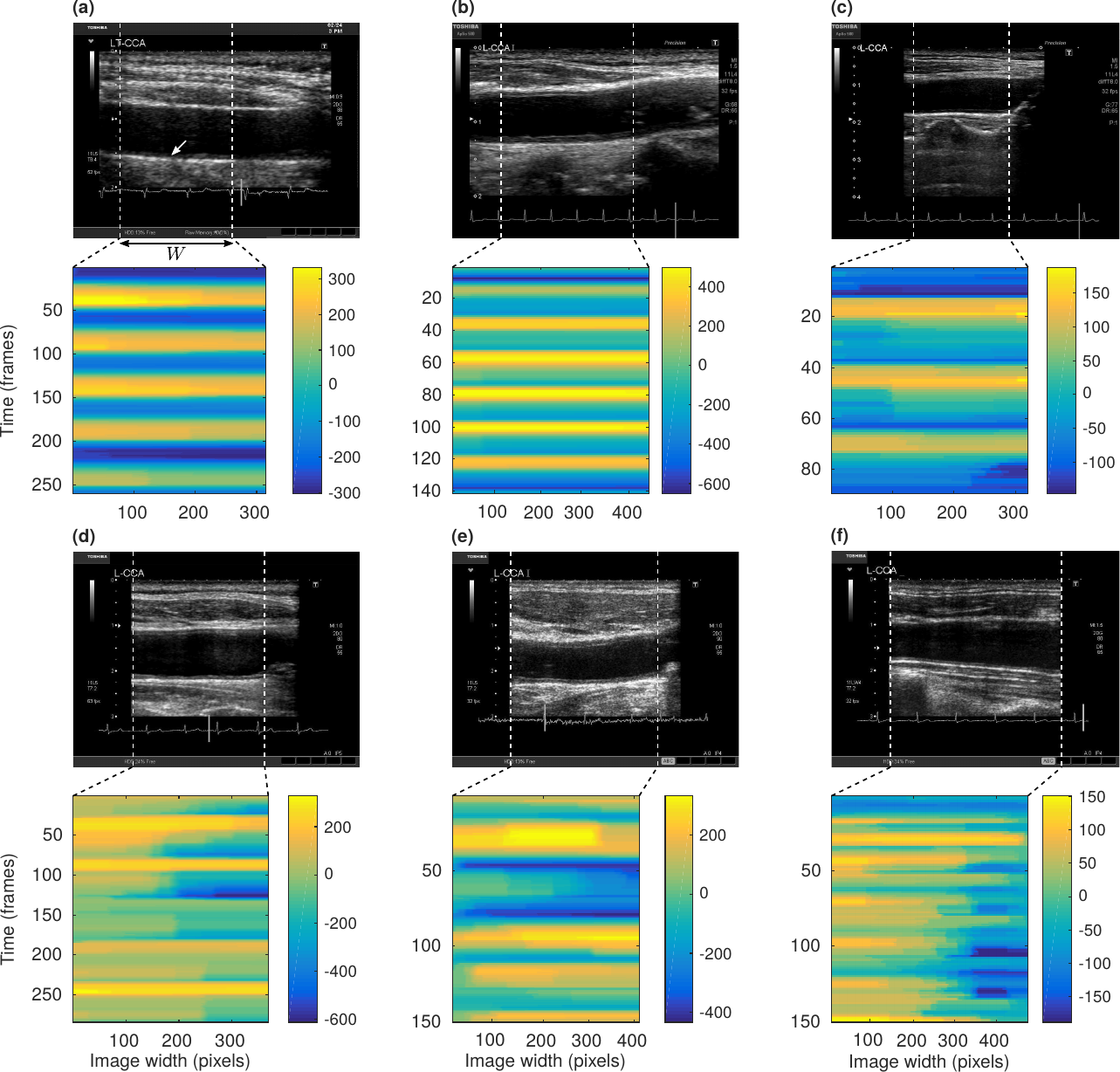}
\caption{\label{fig_maps}
Example result of the longitudinal motion field $\mathcal{X}$ during several cardiac cycles, evaluated in the intima-media complex of the far wall (arrow in panel a) within the region of interest of width $W$ (white dashes), for six different participants of the testing set.
(a-c) The motion field is quite homogeneous.
(d-f) The motion field is rather inhomogeneous, substantial variations are visible in both spatial and temporal directions.
The motion amplitude in $\mu\text{m}$ is indicated by the colorbars.
}
\end{center}
\end{figure}

The very nature of the DBM framework enabled for the extraction of a dense motion field $\mathcal{X}$. Several dynamic patterns, corresponding to different participants, are displayed in Figure~\ref{fig_maps}. Although a quantitative analysis of the tracking accuracy was only performed in three local points per cine-loop, as described above, a qualitative assessment of the motion homogeneity can be obtained from these 2D fields.
\par
The DBM framework was implemented in MATLAB (MATLAB R2016b, The MathWorks Inc., Natick, MA, USA, 2016) and accelerated using Mex/C++, on a desktop computer with a 2.90~GHz processor and 32~Gb RAM. The required computational time to process a typical cine-loop (151 frames, width $W=24.2~\text{mm}$, corresponding to $\Omega=357$ tracked points) was \SI{68}{\second}. More specifically, it was \SI{450}{\milli\second} per frame: \SI{140}{\milli\second} for the layers segmentation, \SI{120}{\milli\second} for the block matching operation, and \SI{190}{\milli\second} for the dynamic programming operation.
\subsection{Intra-participant variability}\label{sec_var}
The intra-participant variability of LOKI with respect to the selection of the $x$-coordinate of the initial measurement point was assessed by means of the $\delta X$ index, as detailed in Section~\ref{sec_stat}.
Among all participants, the mean absolute ($\pm \text{STD}$) value of $\delta X$ was $33~(\pm 48)~$\SI{}{\micro\metre}.
With an average peak-to-peak amplitude $\Delta X$ of  $314~(\pm 163)~$\SI{}{\micro\metre}, the average relative intra-participant variability was $14~(\pm 22)~$\SI{}{\percent} of the total motion amplitude.
\subsection{Motion inhomogeneity}\label{sec_cli}
\begin{figure}[h]
\begin{center}
\includegraphics[width=.5\textwidth]{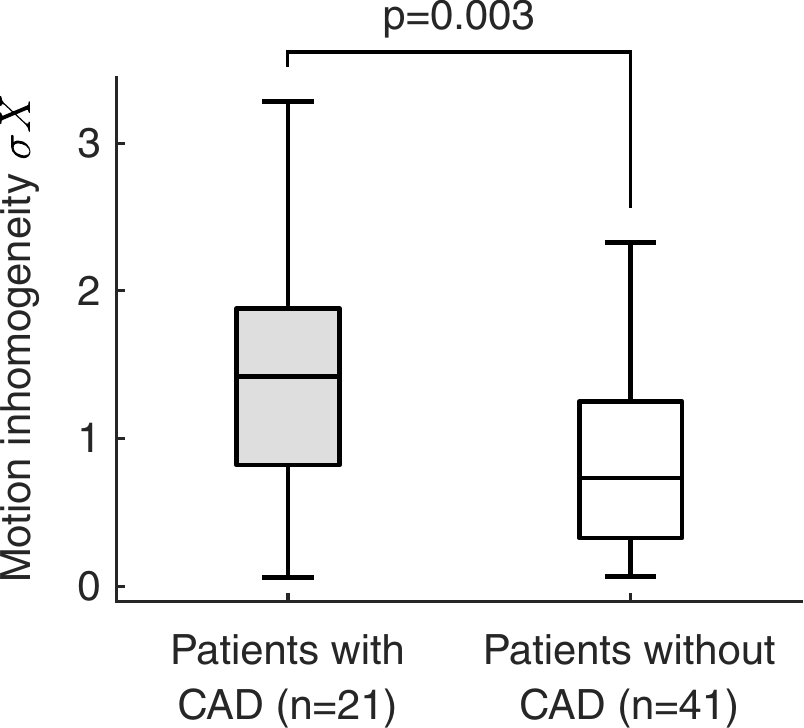}
\caption{\label{fig_stat}
Univariate association between LOKI inhomogeneity index $\sigma X$ and the presence of coronary artery disease (CAD).
Percentiles are indicated by boxes (25th and 75th), inner lines (50th), and error bars (5th and 95th). 
The result of the Mann-Whitney U-test is indicated by the $p$ value.
}
\end{center}
\end{figure}

The association between either the motion inhomogeneity index $\sigma X$ or the peak-to-peak amplitude $\Delta X$ and participants characteristics is provided in Table~\ref{tab_stats}.
The strongest correlates to $\sigma X$ were coronary artery disease ($\beta\text{-coefficient}=.586$, $p=.003$, Figure~\ref{fig_stat}) and systolic blood pressure ($\beta\text{-coefficient}=.015$, $p=.053$).
Interestingly, the association with coronary artery disease remained nearly-significant ($\beta\text{-coefficient}=.484$, $p=.061$) in a multivariable model after adjusting for systolic blood pressure and peak systolic value, the second and third strongest independent correlates.
The peak-to-peak amplitude $\Delta X$ was strongly correlated with peripheral artery disease ($\beta\text{-coefficient}=.165$, $p=.006$) and hypertension ($\beta\text{-coefficient}=.120$, $p=.040$). These results are in accordance with previous findings showing that $\Delta X$ is associated with cardiovascular risk factors~\cite{zahnd2012longitudinal}.
A significant association was also observed between $\sigma X$ and $\Delta X$ ($\beta\text{-coefficient}=-.889$, $p=.001$).
A qualitative analysis of the motion field in all participants did not reveal an association between motion homogeneity and putative local tissue stiffness (\emph{e.g.}~healthy sections or atherosclerotic plaques, Figure~\ref{fig_maps}).
\begin{table}[h!]
\caption{Association of cardiovascular risk factors with LOKI-derived parameters in all 62 participants. The symbols describing the $p$ value are 0 (***) 0.001 (**) 0.01 (*) 0.05 (.) 0.1 (~) 1.}
\label{tab_stats}
\begin{center}
\begin{tabular}{lllllllll}
\hline
& \multicolumn{4}{c}{Inhomogeneity index $\sigma X$} & \multicolumn{4}{c}{Peak-to-peak amplitude $\Delta X$} \\
& \multicolumn{2}{c}{Univariate} & \multicolumn{2}{c}{Multivariable} & \multicolumn{2}{c}{Univariate} & \multicolumn{2}{c}{Multivariable} \\
Parameter & \multicolumn{1}{c}{$\beta$} & \multicolumn{1}{c}{$p$~~~~~} & \multicolumn{1}{c}{$\beta$} & \multicolumn{1}{c}{$p$~~~~~} & \multicolumn{1}{c}{$\beta$} & \multicolumn{1}{c}{$p$~~~~~} & \multicolumn{1}{c}{$\beta$} & \multicolumn{1}{c}{$p$~~~~~}\\ 
\hline
Coronary artery disease   & .586  & .003 (**) & .484      & .061 (.) & .000  & .995      &         &            \\
Hypertension              & .577  & .033 (*)  &             &            & .120  & .040 (*)  & .094  & .151     \\
Systolic blood pressure   & .015  & .053 (.)  & .008      & .357     & -.003 & .151      &         &            \\
Diastolic blood pressure  & .071  & .552      &             &            & -.002 & .532      &         &            \\
Pulse pressure            & .015  & .080 (.)  &             &            & -.002 & .240      &         &            \\
Dyslipidemia              & .496  & .102  &       &     & -.006  & .923      &         &            \\
Peak systolic value       &-.003  & .111      & -.002     & .253     & .000  & .198      &         &            \\
Other heart diseases      & .293  & .131  &       &     & .054  & .196      &         &            \\
Diabetes mellitus         & .206  & .300  &       &     & -.018  & .683      &         &            \\
Gender                    & -.235 & .336      &             &            & .042  & .430      &         &            \\
Age                       & .014  & .372      &             &            & .000 & .898      &         &            \\
Previous stroke           & .162  & .473      &             &            & -.018 & .719      &         &            \\
Intima-media thickness    & -.080 & .520      &             &            & -.002 & .936      &         &            \\
Peripheral artery disease & .154  & .595      &             &            & .165  & .006 (**) & .149  & .012 (*) \\
Estimated glomerular filtration rate & -.003 & .602      &             &            & -.001  & .447      &         &            \\
Total cholesterol         & -.002 & .699      &             &            & .001  & .153      &         &            \\
Glycosylated hemoglobin & -.050 & .725      &             &            & .006  & .820      &         &            \\
Smoking:                  &&&&&&&&\\
- Current                 & -.130 & .773      &             &            & .017  & .857      &         &            \\
- Ex                      & .107  & .770      &             &            & -.072 & .354      &         &            \\
- Never                   & -.134 & .747      &             &            & .006  & .949      &         &            \\
Body mass index           & .006  & .863      &             &            & -.012 & .079 (.)  & -.012 & .069 (.) \\
Other vascular diseases   & .009  & .976      &             &            & .047  & .507      &         &            \\
\hline
\end{tabular}
\end{center}
\end{table}
\section{Discussion}\label{sec_discussion}
The introduced method is devised to simultaneously analyze the motion of a large number of tracked points arranged in a linear fashion. It was applied \emph{in vivo} in carotid B-mode ultrasound cine-loops to extract LOKI, the tissue motion along the vessel axis during the cardiac cycle. The motion is extracted along the entire exploitable length of the imaged artery (as opposed to traditional approaches that only focus on a single point). Such dense motion field analysis was a missing link in LOKI-based studies.
\par
Let us start by briefly recalling the three hypotheses that were investigated in the present study. First, tracking of a single point is more prone to generate erroneous trajectories compared to exploiting the collaboration of multiple points. The increased robustness of DBM was demonstrated in Section~\ref{sec_rob} against a state-of-the-art method based on single-point tracking, with an average absolute error of $150 \pm$\SI{163}{\micro\metre} compared to $191 \pm$\SI{270}{\micro\metre}. Second, the selection of a single point is an analyst-dependent procedure and may hinder the reliability of results exploitation. The impact caused by single-point selection when measuring the peak-to-peak motion amplitude was assessed in Section~\ref{sec_var}, and showed that the intra-participant variability $\delta X$ was $33 \pm$\SI{48}{\micro\metre}. Although relatively reduced compared to the total magnitude of the motion, this additional source of error can be avoided with a global analysis. Third, motion trajectories resulting from a single point do not allow the analysis of the displacement field (in)homogeneity across the entire length of the artery, which may reflect cardiovascular health. The results presented in Section~\ref{sec_cli} establish that the LOKI-derived inhomogeneity index $\sigma X$ is strongly associated with coronary heart disease ($\beta\text{-coefficient}=.586$, $p=.003$, Figure~\ref{fig_stat}).
\par
The collaboration of all the points allowed for improved robustness. The spatial consistency of the motion was implicitly exploited within the front propagation step (Eq.~\ref{eq_propag}), since the global optimum was preferred over a series of independent best matches, thus reducing the risk of individual tracking errors. To phrase it differently, the set of conditions ruling the behaviour between neighboring points could successfully maintain the cohesion of the mesh and prevent sudden jumps or cumulative drift errors. On the other hand, the mesh was flexible enough for the points to follow the motion of the tissues over time. This phenomenon is illustrated in Figure~\ref{fig_traj}c. At this point, it is useful to differentiate accuracy and robustness. While accuracy describes the ability of a method to yield results close to the ground truth, robustness represents the capacity of a method to avoid failure and provide an exploitable result. In the present situation, it is unlikely that the DBM method could have a positive impact on the tracking accuracy \emph{per se}, however the robustness was improved. Moreover, it should be noted that the combinatorial analysis, solved by means of dynamic programming, also enabled a simultaneous motion computation for a large collection of points in a rather limited time due to the intrinsic reduction of the solution space.
\par
The motion along the $y$ direction was guided by the contour segmentation and a reduced search window, whereas the motion along the $x$ direction, substantially more challenging to assess, was obtained by means of a dynamic programming approach. The user interaction was minimal, only the left and right borders of the ROI to be processed had to be selected to discard non-exploitable regions of the image. In addition to providing motion information for the entire ROI, this procedure (namely, region selection rather than point selection) reduces user variability compared to individual point selection and facilitates full-automation.
\par
A fair evaluation was conducted by using a training set ($n=20$) to define the optimal parameter settings and an independent testing set ($n=42$) to quantify the accuracy of the method. Results demonstrated that the DBM framework was able to perform in the testing set almost as well as in the training set, showing the ability of the method to analyze previously unseen data. It is also noteworthy that the method was evaluated in challenging conditions, as image quality was overall quite poor, mostly for two reasons. First, images acquired with the ultrasound scanner were rather coarse (pixel size of \SI{70}{\micro\metre}). Second, the involved participants were all elderly patients suffering from atherosclerosis, and therefore data was more challenging to process than images from young healthy volunteers due to a less favorable echogenicity.
\par
Although the motion inhomogeneity index $\sigma X$ is amplitude-independent since it is derived from the normalized motion field, this index was correlated to the peak-to-peak amplitude $\Delta X$ ($\beta\text{-coefficient}=-.889$, $p=.001$). Nevertheless, confronting those parameters to the participants characteristics indicated that $\sigma X$ and $\Delta X$ are strongly correlated to different predictor variables (Table~\ref{tab_stats}). These preliminary findings suggest that $\sigma X$ do not simply replicate the information already captured by $\Delta X$ but instead represents a complementary measure of cardiovascular risk. A comprehensive analysis of the association of motion-derived parameters with risk factors was beyond the scope of the present study. Future work involving a greater number of participants will be conducted to confirm the clinical usefulness of analyzing the spatial inhomogeneity of the motion along the wall to characterize at-risk individuals.
\section{Conclusion}\label{sec_conclusion}
The main contribution of this article is the introduction of a novel motion estimation framework, called DBM for Dynamic Block Matching, specifically designed to extract the temporal trajectory of a large collection of points along a direction of interest. The method was applied in ultrasound cine-loops of the common carotid artery to analyze the motion of the intima-media complex in the direction parallel to the blood flow (\emph{i.e.},~longitudinal kinetics, LOKI) and extract a dense motion field. Evaluated \emph{in vivo} in 42 participants, the proposed method demonstrated good tracking performances compared to manually established references. The analysis of the inhomogeneity of the tissue motion across the entire length of the artery showed a strong association with the presence of coronary artery disease. These findings suggest that the proposed method may provide complementary information for cardiovascular risk assessment.
\section*{Conflict of interest}
No benefits in any form have been or will be received from a commercial party related directly or indirectly to the subject of this manuscript.
\begin{acknowledgements}
This work was partly supported by the JSPS \#PE16208 funding as well as the MEXT KAKANHI \#26108004 funding.
\end{acknowledgements}

\end{document}